\documentclass[journal]{IEEEtran}
\usepackage{amsmath,amsfonts}
\usepackage{algorithmic}
\usepackage{algorithm}
\usepackage{array}
\usepackage[caption=false,font=normalsize,labelfont=sf,textfont=sf]{subfig}
\usepackage{textcomp}
\usepackage{stfloats}
\usepackage{url}
\usepackage{verbatim}
\usepackage{graphicx}
\usepackage{siunitx}
\usepackage{float}
\usepackage{tikz}
\begin{document}

\title{Analytical Phasor-Based Fault Location Enhancement for Wind Farm Collector Networks}

\author{Alailton~J.~Alves~Junior, Daniel~Barbosa, Ricardo A. S. Fernandes, and Denis~V.~Coury%
\thanks{A. J. Alves Júnior, R. A. S. Fernandes and D. V. Coury are with the Department of Electrical and Computer Engineering, São Carlos School of Engineering, University of São Paulo, São Carlos, SP 13566-590, Brazil (e-mail: alailtonjunior@usp.br; ricardo.asf@usp.br; coury@sc.usp.br).}%
\thanks{D. Barbosa is with the Department of Electrical Engineering, Federal University of Bahia, Salvador, BA 40170-110, Brazil (e-mail: dbarbosa@ufba.br).}%
\thanks{(Corresponding author: Alailton J. Alves Junior.)}
}

% The paper headers
\markboth{Journal of \LaTeX\ Class Files,~Vol.~14, No.~8, August~2021}%
{Shell \MakeLowercase{\textit{et al.}}: Analytical Phasor-Based Fault Location Enhancement for Wind Farm Collector Networks}

\IEEEpubid{0000--0000/00\$00.00~\copyright~2021 IEEE}

\maketitle

% IEEE preprint notice for arXiv
\begin{center}
  \small
  This work has been submitted to the IEEE for possible publication. 
  Copyright may be transferred without notice, after which this version 
  may no longer be accessible.
\end{center}

\begin{abstract}
The increasing integration of Inverter-Based Resources (IBRs) is reshaping fault current characteristics, presenting significant challenges to traditional protection and fault location methods. This paper addresses a key limitation in fault location within wind farm collector networks, i.e., one-terminal phasor-based methods become inaccurate when IBRs are electrically located downstream from the fault. In such cases, the voltage drop caused by IBR fault current injections is not captured by the Intelligent Electronic Device (IED), resulting in a systematic overestimation of fault distance. To mitigate this issue, a general compensation framework was proposed by augmenting classical loop formulations with a distance-dependent voltage correction term. The methodology was derived analytically using a sequence-domain representation and generalized to multiple fault types through a unified notation. It maintains the simplicity and interpretability of conventional approaches and can be implemented using only local measurements. The method was evaluated through EMT simulations in PSCAD using a realistic wind farm model. Results show significant improvements in location accuracy, with average and maximum errors notably reduced, especially for ground-involved faults where reductions exceed 90\%. Furthermore, the compensation eliminates sensitivity to wind penetration levels and ensures uniform performance across feeders, positioning the method as a practical solution for modern renewable-dominated grids.
\end{abstract}

\begin{IEEEkeywords}
Fault location, inverter-based resources, wind farm collector, one-terminal methods
\end{IEEEkeywords}

\section{Introduction}

\IEEEPARstart{G}{lobal} commitments to decarbonization, supported by national targets, market instruments, and sustained cost declines, have accelerated the deployment of wind and solar generation worldwide~\cite{lee2024global}. As the resource mix shifts from synchronous machines to converter-interfaced generation, short-circuit characteristics are being reshaped across transmission, distribution, and plant-level networks: fault magnitudes, temporal waveforms, and sequence components increasingly diverge from the assumptions that historically guided protection engineering and fault-analysis practice~\cite{IEEE2018}.

These shifts are driven chiefly by the dynamics of Inverter-Based Resources (IBRs). Unlike synchronous machines, IBRs inject fault current through control-driven mechanisms that inherently limit magnitude, shape the current waveform via fast control loops, and alter sequence components~\cite{transmission_chowdhury_2021, impact_haddadi_2021}. Therefore, the accurate fault location is a prerequisite for dependable system operations, as it shortens isolation and restoration times, and limits service disruptions. Accordingly, this study examines the performance of traditional fault location methods in IBR-rich networks and proposes a method tailored to their specific characteristics.

Most fault-location methods are categorized into four groups: phasor-based, voltage-sag, traveling-wave, and data-driven. Among these, phasor-based methods are often preferred because they are straightforward to implement, interpretable, and compatible with existing measurement infrastructures~\cite{DAS2014}. 

Within phasor-based methods, single-ended schemes rely exclusively on local measurements to estimate the fault distance. Common methods include the impedance-based formulation~\cite{ZIEGLER2011}, reactance-based estimators~\cite{CAPAR2014}, and the family of Takagi-type solutions~\cite{TAKAGI1982,DAS2014,SEL2018}. Two-ended schemes extend this approach by incorporating synchronized measurements at both line terminals, which requires a communication channel; notable contributions include~\cite{GIRGIS1992,JJ1990,PR2011,HE2011}. Despite this variety, phasor-based methodologies are grounded in standard short-circuit assumptions associated with synchronous-machine behavior.

Motivated by these limitations, recent literature has begun to explicitly account for IBR effects in fault-location algorithms. At the distribution and microgrid scales, three studies exemplify these directions~\cite{Kahnamouei2023OptimizedFLIIDG, Apostolopoulos2022Unsync, Matthews2019FaultCurrentCorrection}. In \cite{Kahnamouei2023OptimizedFLIIDG}, the authors proposed an optimization-based three-phase fault-location method that fuses synchronized and unsynchronized measurements, models IBR fault response without Phasor Measurement Unit data, and explicitly treats load-value uncertainty. Additionally, in \cite{Apostolopoulos2022Unsync}, the authors rely only on unsynchronized source-side phasors, avoid IBR source-impedance modeling, and jointly estimate fault distance, resistance, and synchronization angles. A supercapacitor fault current correction was proposed by \cite{Matthews2019FaultCurrentCorrection}, enabling conventional overcurrent protection and fault location to remain effective under inverter-limited currents.

In transmission systems with high IBR penetration, two complementary research directions have emerged. For instance, in \cite{Davi2023IBRMultiMethod} the authors assessed ten impedance-based fault-location schemes, comprising both one- and two-terminal methods, using several PSCAD simulations covering a wide range of IBR control strategies and short-circuit levels. Based on these results, they proposed a multi-method selection scheme that significantly reduces the location error when no communication channel is available.

The second direction focuses on sequence-aware single-ended formulations tailored to the converter behavior. In \cite{Mobashsher2024LocalIBR}, the authors developed a sequence-network model for IBRs showing that the apparent negative-sequence impedance became effectively infinite when grid codes did not mandate negative-sequence injection, and designed a single-ended locator that employed negative- and zero-sequence polarizing currents, incorporated remote-infeed and line-charging effects, and maintained accuracy across wide ranges of fault resistance and grid-code settings.

Despite this progress, the collector system within wind farms remains comparatively underexplored. Collector feeders are typically radial and host multiple IBRs at varying electrical distances from a prospective fault. Addressing this gap, the authors of ~\cite{DAVI2025MULTI} investigate the performance of non-conventional single-ended phasor-based methods in a wind-farm collector system. The study highlights the most promising traditional methods for each fault condition, but does not account for IBR influence on these methods.

In order to address these limitations, this paper presents an enhanced single-ended, phasor-based fault-location method specifically designed for wind-farm collector systems. The key contribution is a transparent modification of the classical loop equation that explicitly incorporates the voltage drops caused by intermediate IBR-injected currents. By doing so, the proposed approach enables high-accuracy fault location in renewable-rich environments while preserving the simplicity and local-measurement dependence of conventional schemes.

The paper proceeds as follows. Section~\ref{sec:problem} analyzes how IBR placement with respect to the fault point affects the single-ended loop and motivates the inclusion of a compensation term in the loop equation. Section~\ref{sec:methodology} develops the compensated formulation, presents the full Single Line-to-Ground (SLG) derivation, and generalizes the notation for other fault types. Section~\ref{sec:test-system} describes the test system and scenarios. Section~\ref{sec:results} evaluates the proposed method's performance against a classical phasor-based scheme. Finally, Section~\ref{sec:conclusion} summarizes the findings.

\section{IBR Effect on One-Terminal Fault Locators}
\label{sec:problem}

\begin{figure}[!h]
    \centering
    \includegraphics[width=\linewidth]{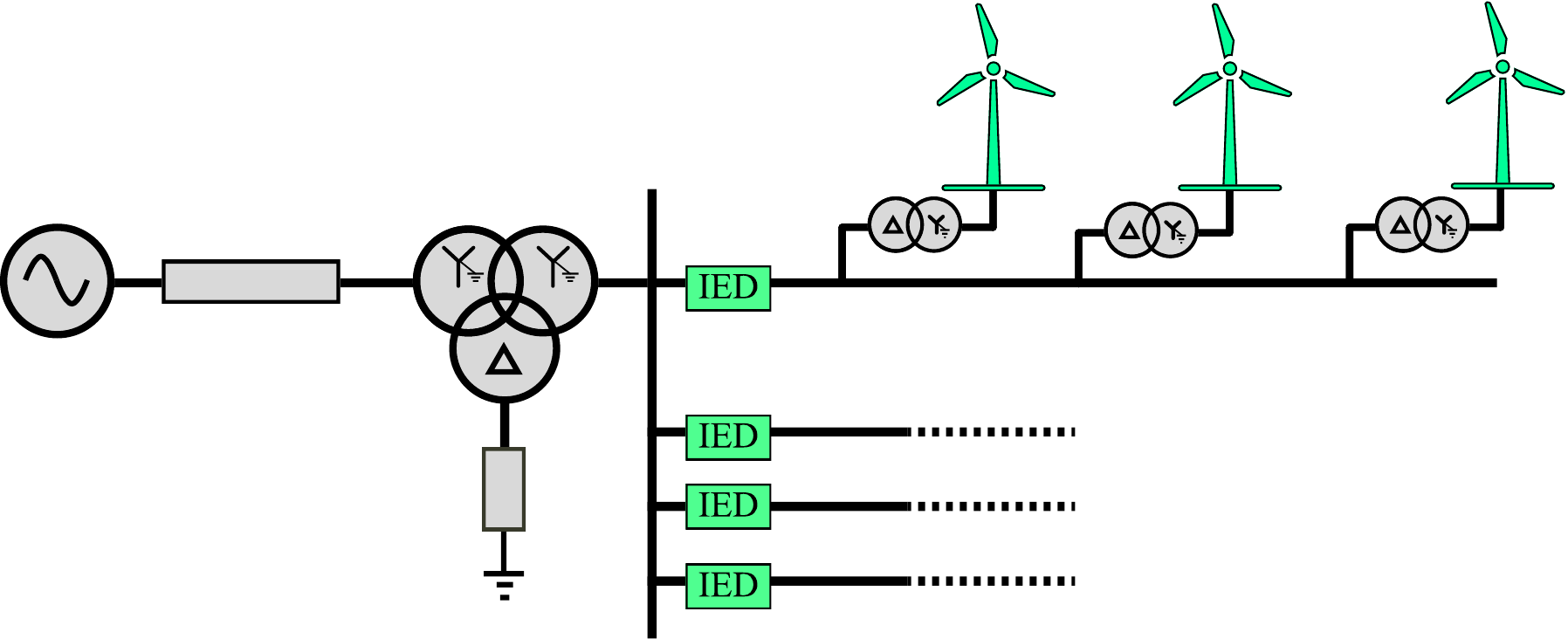}
    \vspace{-0.7cm}
    \caption{Schematic overview of a wind-farm collector system with multiple IBRs connected along radial feeders.}
    \label{fig:test-system-intro}
\end{figure}

Modern wind farms comprise numerous wind turbine generators (WTGs) distributed across multiple radial feeders that converge at a central collector substation. This collector system is responsible for aggregating the generated power and exporting it to the transmission network, as shown in Fig.~\ref{fig:test-system-intro}. Each feeder consists of a primary distribution line and several lateral branches, with IBRs interfacing the WTGs to the grid.
\vspace{-12pt}

During fault conditions, IBRs inject positive-sequence current with a limited amplitude and controlled characteristics, fundamentally different from the behavior of traditional synchronous generators. While this distinction in current profile is significant, the most critical factor in fault location analysis is not the nature of the injected current itself, but rather the position of the IBR in relation to the fault point and the Intelligent Electronic Device (IED) installed at the feeder's extremity. The presence of an IBR between the IED and the fault introduces a substantial change in the fault loop, which can distort the voltage and current measurements used for distance estimation.

Traditional phasor-based fault location algorithms rely on the premise that the current flowing through the IED is the same current traversing the segment of the line between the IED and the fault. Under this assumption, the ratio between the voltage and current phasors, termed the loop impedance, reflects the impedance of the line segment up to the fault point \cite{SEL2018}. However, this assumption fails in the presence of IBRs depending on their location relative to the fault.

\subsection{Fault Located Upstream of the IBR}

\begin{figure}[!t]
    \centering
    \includegraphics[width=1\linewidth]{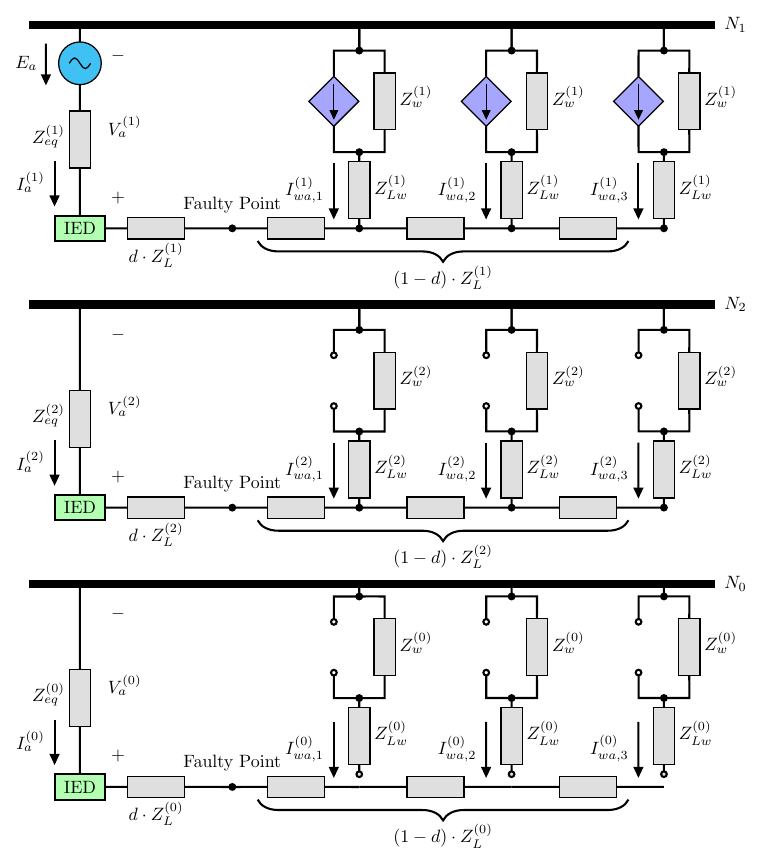}
    \vspace{-0.8cm}
    \caption{Sequence-network representation of a feeder with all IBRs located upstream from the fault.}
    \label{fig:seq-network}
    \vspace{-14pt}
\end{figure}

In Fig.~\ref{fig:seq-network}, the feeder was configured in per-unit distance from the IED at the local end to the remote end. A fault at location $d\in[0,1]$ was measured from the IED toward the remote end, and the $k$-th IRB was located at $d_{w,k}\in[0,1]$. Under the scenario, the inverters lie electrically beyond the fault, so $d<d_{w,k}$ for every $k$. Let $N$ denote the number of inverters, with locations ${d_{w,k}}|_{k=1}^{N}$. The total positive-sequence line impedance was $Z_L^{(1)}$. Therefore, a per-unit segment of length $\Delta d$ has impedance $\Delta d,Z_L^{(1)}$. Analogous definitions apply to the negative- and zero-sequence impedances, $Z_L^{(2)}$ and $Z_L^{(0)}$. The current injected by the $k$-th inverter was denoted $I_{wa,k}$.

In this configuration, the entire voltage drop between the IED and the fault point was captured by the IED measurements. Although a portion of the fault current injected by the IBR may flow toward the grid and pass through the IED, the relation between voltage and current along the line segment from the IED to the fault remains unaltered. 

This occurred because the entire voltage drop in the faulty segment was captured in the measured voltage, and the associated current represents the net current through that same segment. As a result, the loop impedance computed from the phasors measured at the IED still reflects the impedance of the line up to the fault. Consequently, the conventional loop formulation remains valid in this scenario and can be expressed in \eqref{eq:loop_traditional}:

\begin{equation}
    Z_{\text{loop}} = \frac{V_{\text{loop}}}{I_{\text{loop}}},
    \label{eq:loop_traditional}
\end{equation}

\noindent where $V_{\text{loop}}$ and $I_{\text{loop}}$ are the phasor voltage and current measured locally at the IED. This formulation provides an accurate representation of the impedance between the IED and the fault point, assuming no intermediate IBRs alter the current path.

\subsection{Fault Located Downstream from the IBR}

\begin{figure}[!ht]
    \centering
    \includegraphics[width=1\linewidth]{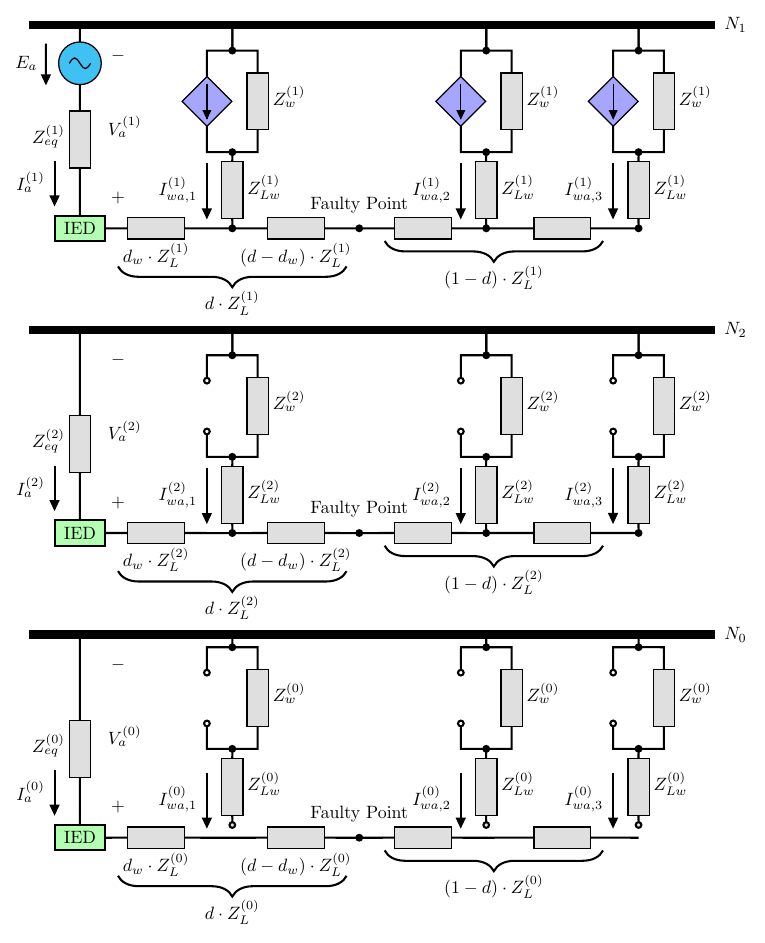}
    \vspace{-0.8cm}
    \caption{Sequence-network representation of a feeder with one IBR located downstream from the fault.}\label{fig:seq-network-downstream}
    \vspace{-14pt}
\end{figure}

In contrast, when the fault was located downstream from at least one IBR illustrated in Fig.~\ref{fig:seq-network-downstream}, the assumption that the fault current measured at the IED traverses the entire faulted segment was no longer valid. This occurs when there exists at least one $k$ such that $d_{w,k} < d$, meaning the IBR was positioned upstream of the fault.

In such configurations, a portion of the fault positive-sequence current injected by the IBR must flow through part of the line between the IBR and the fault point. This segment, denoted by $(d - d_{w,k})\,Z_L^{(1)}$, introduces a voltage drop that was not visible to the IED, as the IED does not measure the IBR's injected current directly. Therefore, while the IED records the voltage drop only up to the IBR, it does not capture the additional drop across the remainder of the line up to the fault. This leads to a mismatch between the measured voltage and the actual current path, distorting the loop impedance and resulting in an overestimation of the fault distance.

To correct this discrepancy, a compensation term was added to the loop voltage to reconstruct the total voltage drop across the actual faulted segment. This leads to a modified expression for the loop impedance that includes the IBR-induced voltage component, as shown in \eqref{eq:loop_compensated}:

\begin{equation}
    Z_{\text{loop}} = \frac{V_{\text{loop}} + V_{\text{comp}}(d)}{I_{\text{loop}}},
    \label{eq:loop_compensated}
\end{equation}

\noindent where $V_{\text{comp}}(d)$ represents the voltage drop introduced by the IBR current flowing through the segment between the IBR and the fault. By incorporating this compensation term, the fault location algorithm can restore the accuracy lost due to the distortion caused by upstream IBRs.

This correction becomes essential in wind farm collector systems, where multiple IBRs are distributed along the feeders.

\section{Methodology}
\label{sec:methodology}

To address the challenge posed in Section~\ref{sec:problem}, the present section develops a transparent and fully traceable compensation framework that augments a one-terminal phasor-based fault-location method. The core idea was to include a distance-dependent voltage term, $V_{\text{comp}}(d)$, in the loop formulation to correct the effect of inverter injections that do not traverse the entire path between the IED and the fault. %As $V_{\text{comp}}(d)$ depends on the unknown distance $d$, the estimate was obtained iteratively. % as summarized in Figure~\ref{fig:interactive}.

\subsection{Single-Line-to-Ground Fault Derivation}
\label{sec:methodology-wind}

\begin{figure}[!ht]
  \centering
  \includegraphics[width=1\linewidth]{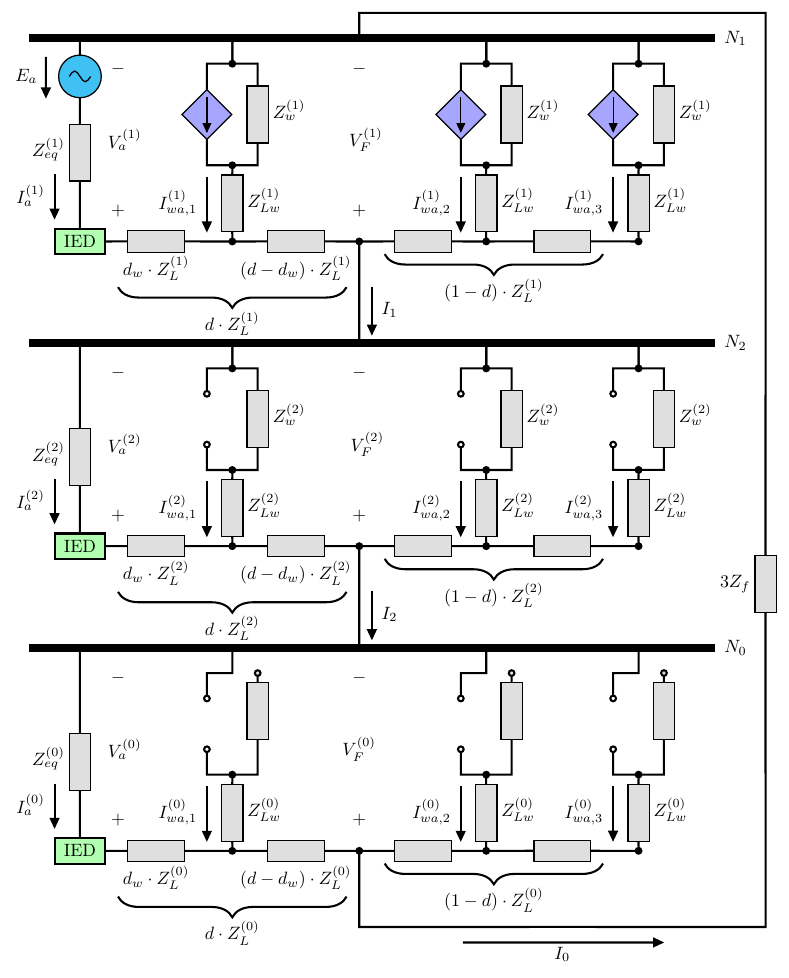}
  \vspace{-0.8cm}
  \caption{Sequence representation of an SLG fault on a feeder with intermediate inverters.}
  \label{fig:seq-network-SLG}
  \vspace{-12pt}
\end{figure}

The formulation was first obtained for the SLG fault case and then generalized. Fig.~\ref{fig:seq-network-SLG} illustrates the interconnection used to represent the series connection of the three sequence networks during an SLG fault.

Within this setting, consider the faulted phase $a$ located at distance $d$ from the IED, and its currents and voltages measurements are represented by $I_a$, $I_b$, $I_c$ and $V_a$, $V_b$, $V_c$ respectively. 

%The total positive-sequence line impedance is denoted by $Z_L^{(1)}$, thus the impedance of a per-unit segment $\Delta d$ equals $\Delta d\,Z_L^{(1)}$; analogous definitions hold for $Z_L^{(2)}$ and $Z_L^{(0)}$. Within this setting, consider the faulted phase $a$ located at distance $d$ from the IED. Let $N$ be the number of inverters with locations $d_{w,k}\|_{k=1}^{N}$. And, The IED current and voltage measurements are represented by $I_a$, $I_b$, $I_c$ and $V_a$, $V_b$, $V_c$ respectively. And the positive-sequence current injected by the $k$-th inverter is denoted by $I_{wa,k}^{(1)}$.

By extracting the loop relations in Fig.~\ref{fig:seq-network-SLG}, the voltage at the fault node was given by Equation~\eqref{eq:PG_fault_voltage}:

\begin{equation}
  V_F \;=\; V_F^{(1)} + V_F^{(2)} + V_F^{(0)},
  \label{eq:PG_fault_voltage}
\end{equation}

\noindent where $V_F$ represents the voltage at the fault point, and the upscript $(\cdot)^{(1)}$, $(\cdot)^{(2)}$, and $(\cdot)^{(0)}$ denote the positive-~, negative-, and zero-sequence components, respectively.

\begin{figure}[!ht]
\centering\includegraphics[width=0.95\linewidth]{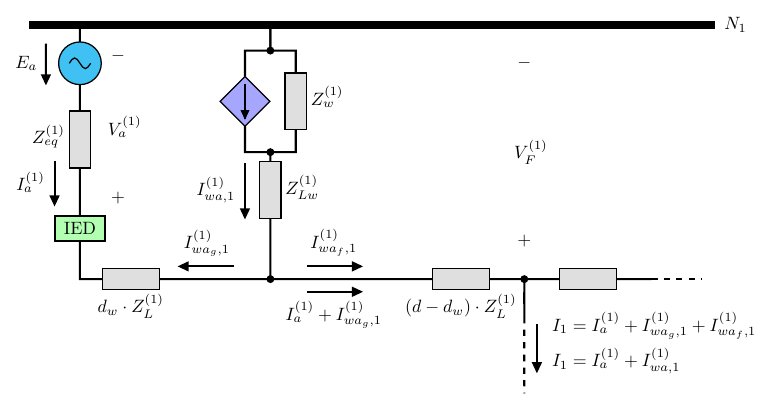}
  \vspace{-0.5cm}
  \caption{Current-divider illustration for one IBR in the positive-sequence network.}
  \label{fig:ibr-contribution}
  \vspace{-8pt}
\end{figure}

Firstly, analysing the positive-sequence network, each IBR injects a current $I_{wa,k}^{(1)}$ that splits between an upstream branch toward the IED and a downstream branch toward the fault, as depicted in Fig.~\ref{fig:ibr-contribution}. Denoting these components by $I_{wa_{g},k}^{(1)}$ (to the grid/IED side) and $I_{wa_{f},k}^{(1)}$ (to the fault side), the partition is shown in \eqref{eq:wind_current_components}:

\begin{equation}
 I_{wa,k}^{(1)} \;=\; I_{wa_{g},k}^{(1)} + I_{wa_{f},k}^{(1)}.
 \label{eq:wind_current_components}
\end{equation}

Consequently, the positive-sequence current seen by the IED equals the grid-side contribution minus the portion of the IBR injections flowing back toward the IED, estimated by \eqref{eq:3P_1}:

\begin{equation}
I_a^{(1)} \;=\; I_{ga}^{(1)} \;-\; \sum_{k=1}^{N} I_{wa_{g},k}^{(1)},
\label{eq:3P_1}
\end{equation}

\noindent where $I_{ga}^{(1)}$ denotes the grid-side contribution observed at the IED. Equation~\eqref{eq:3P_1} makes clear that the IED does not necessarily measure the entire inverter injection. For any IBR with $d_{w,k}<d$, the current flowing toward the fault increases downstream from $d_{w,k}$. Consequently, the segment of length $(d-d_{w,k})$ experiences an additional voltage drop $(d-d_{w,k})\,Z_L^{(1)}\,I_{wa,k}^{(1)}$ that is not captured at the IED.

To expose the role of downstream inverter injections, consider a single inverter at $d_{w,1}$ with $d_{w,1}<d$ just as illustrated in Fig.~\ref{fig:seq-network-SLG}. Splitting the line into IED$\to$inverter (length $d_{w,1}$) and inverter$\to$fault (length $d-d_{w,1}$) yields the positive-sequence voltage drops in Equations~\eqref{eq:wind_current_components1} and~\eqref{eq:wind_current_components2}:

\begin{align}
  V_{F}^{(1)} \;&=\; V_a^{(1)} - \bigl(V_{\text{drop},1} + V_{\text{drop},2}\bigr) \nonumber\\[1mm]
  V_{\text{drop},1} \;&=\; I_a^{(1)} \, d_{w,1} \, Z_L^{(1)}
  \label{eq:wind_current_components1}\\
  V_{\text{drop},2} \;&=\; \Bigl[I_{wa_{f},1}^{(1)} + \bigl(I_a^{(1)} + I_{wa_{g},1}^{(1)}\bigr)\Bigr] (d-d_{w,1}) \, Z_L^{(1)}
  \label{eq:wind_current_components2}
\end{align}

\noindent being $V_{\text{drop},1}$ the voltage drop along the upstream segment, while $V_{\text{drop},2}$ is the drop along the downstream segment. Combining Equations~\eqref{eq:wind_current_components1} and~\eqref{eq:wind_current_components2} leads to \eqref{eq:VF1_combined}:

\begin{align}
  V_{F}^{(1)} \;&=\; V_a^{(1)} \;-\; I_a^{(1)} \, d \, Z_L^{(1)}\nonumber\\
  &\quad \quad - \bigl(I_{wa_{f},1}^{(1)} + I_{wa_{g},1}^{(1)}\bigr) (d-d_{w,1}) \, Z_L^{(1)}, \nonumber\\
               \;&=\; V_a^{(1)} \;-\; I_a^{(1)} \, d \, Z_L^{(1)} \;-\; I_{wa,1}^{(1)} (d-d_{w,1}) \, Z_L^{(1)}.
  \label{eq:VF1_combined}
\end{align}

It was possible to observe that Equation~\eqref{eq:VF1_combined} makes the decomposition explicit. The conventional loop $V_a^{(1)} - I_a^{(1)} d\, Z_L^{(1)}$ applies when no IBR was located between the IED and the fault, whereas any IBR at $d_{w,k}<d$ introduces an additional voltage drop over the segment of length $(d-d_{w,k})$ because its injected current does not traverse the entire distance $d$. By superposition, for $N$ inverters, the general positive-sequence fault voltage is given by Equation~\eqref{eq:3P_2}:

\begin{equation}
  \begin{split}
    V_{F}^{(1)} &\;=\; V_a^{(1)} \;-\; I_a^{(1)} \, d \, Z_L^{(1)} \\& \quad\;-\;\sum_{k=1}^{N}
  \begin{cases}
 (d - d_{w,k}) \, Z_L^{(1)} \, I_{wa,k}^{(1)}, & \text{if } d_{w,k} < d, \\
    0, & \text{otherwise,}
  \end{cases}
  \label{eq:3P_2}
  \end{split}
\end{equation}
which motivates the definition of the compensation term in \eqref{eq:compensation_voltage}.
\begin{equation}
  V_{\text{comp}}^{(SLG)}(d) \;=\; -\sum_{k=1}^{N} 
  \begin{cases}
 (d - d_{w,k}) \, Z_L^{(1)} \, I_{wa,k}^{(1)}, & \text{if } d_{w,k} < d, \\
    0, & \text{otherwise.}
  \end{cases}
  \label{eq:compensation_voltage}
\end{equation}

Consistent with \cite{en14041050}, typical inverter controls suppress negative- and zero-sequence injections. Under this assumption, the negative- and zero-sequence fault-point voltages contain no inverter-current terms and are written in Equations~\eqref{eq:3P_4} and~\eqref{eq:3P_5}:

\begin{align}
  V_{F}^{(2)} \;&=\; V_a^{(2)} \;-\; I_a^{(2)} \, d \, Z_L^{(2)}
  \label{eq:3P_4}\\
  V_{F}^{(0)} \;&=\; V_a^{(0)} \;-\; I_a^{(0)} \, d \, Z_L^{(0)}.
  \label{eq:3P_5}
\end{align}

Together with the positive-sequence relation in Equation~\eqref{eq:3P_2}, these expressions fully specify the sequence-domain fault voltages ($V_{F}^{(1)}, V_{F}^{(2)}, V_{F}^{(0)}$). For other fault types, the per-sequence voltage relations themselves remain unchanged; what differs is the interconnection of the sequence networks at the fault. For example, in an SLG fault, the positive-, negative-, and zero-sequence networks are connected in series, whereas in a line-to-line fault, the positive- and negative-sequence networks are in series while the zero-sequence branch is open.

Substituting Equations~\eqref{eq:3P_2},~\eqref{eq:3P_4}, and~\eqref{eq:3P_5} into the fault constraint, Equation~\eqref{eq:PG_fault_voltage} yields the fault voltage loop in Equation~\eqref{eq:PG_4}.

\begin{align}
  V_F \;&=\; V_a^{(1)} + V_a^{(2)} + V_a^{(0)} + V_{\text{comp}}^{(SLG)}(d)\nonumber\\
          & \qquad\;-\; d\!\left(Z_L^{(1)} I_a^{(1)} + Z_L^{(2)} I_a^{(2)} + Z_L^{(0)} I_a^{(0)}\right), \nonumber\\
    I_F R_F\;&=\;\; V_a \;-\; d \, Z_L^{(1)} \bigl(I_a + K_0 I_a^{(0)}\bigr) \;+\; V_{\text{comp}}^{(SLG)}(d).
  \label{eq:PG_4}
\end{align}

\noindent where $K_0 = Z_L^{(0)}/Z_L^{(1)}$ is the zero-sequence compensation factor, $I_F$ and $R_F$ accounts for the fault current and resistance respective. The negative-sequence line follows under the explicit assumption $Z_L^{(2)} \approx Z_L^{(1)}$. 

Rearranging Equation~\eqref{eq:PG_4}, it was possible to isolate the loop-impedance contribution, leading to Equation~\eqref{eq:PG_final2}:

\begin{equation}
  Z_{\text{loop}} \;=\; d \, Z_L^{(1)} \;=\; \frac{V_a + V_{\text{comp}}^{(SLG)}(d) - I_F \,R_F}{\,I_a + K_0 I_a^{(0)}\,},
  \label{eq:PG_final2}
\end{equation}

\noindent which holds the same structure as the classical loop equation but includes the compensation voltage $V_{\text{comp}}(d)$. Therefore, this formulation can be extended to any phasor-based fault-location method or even distance protection schemes.

\subsection{Extension to Other Fault Types}

The proposed formulation can be expanded to consider the general loop equation for any fault type, as shown in \eqref{eq:PG_final}:

\begin{equation}
  Z_{\text{loop}} \;=\; d \, Z_L^{(1)} \;=\; \frac{V_{\text{loop}} + V_{\text{comp}}(d) - I_F \,R_F}{\,I_{\text{loop}}\,},
  \label{eq:PG_final}
\end{equation}

\noindent where, $V_{\text{loop}}$ and $I_{\text{loop}}$ are the voltage and current loop for the fault type under consideration.

For distance estimation, isolating the $d$ term and applying the imaginary part operator to remove the fault resistance, leads to \eqref{eq:distance_final}:

\begin{equation}
  d \;=\; \frac{\text{Im}\{\left (V_{\text{loop}} + V_{\text{comp}}(d)\right ) \cdot I_F^* \}}{\text{Im}\{Z_L^{(1)} \cdot I_{\text{loop}} \cdot I_F^* \}}
  \label{eq:distance_final}
\end{equation}

\noindent being $\text{Im}\{\cdot\}$ the imaginary part operator and $I_F^*$ is the complex conjugate of the fault current. $V_{\text{comp}}(d)$ is the compensation voltage for any fault type given by Equation~\eqref{eq:compensation_voltage_generic2}:

\begin{equation}
  V_{\text{comp}}(d) \;=\; -\sum_{k=1}^{N} 
  \begin{cases}
 (d - d_{w,k}) \, Z_L^{(1)} \, I_{w,k}, & \text{if } d_{w,k} < d, \\
    0, & \text{otherwise.}
  \end{cases}
  \label{eq:compensation_voltage_generic2}
\end{equation}

\noindent where $I_{w,k}$ is the sequence-appropriate inverter current for the loop under consideration. 

Table~\ref{tab:compensation} consolidates representative loops for common faults, explicitly linking each row to Equations~\eqref{eq:distance_final} and \eqref{eq:compensation_voltage_generic2}. Rows marked ($^{*}$) follow loop formulations adopted in~\cite{Davi2023IBRMultiMethod}.

\begin{table}[!ht]
  \renewcommand{\arraystretch}{1.3}
  \centering
  \caption{Loop quantities for different fault types.}
  \label{tab:compensation}
  \begin{tabular}{cccc}
    \hline
    \textbf{Fault Type} & \textbf{$V_{\text{loop}}$} & \textbf{$I_{\text{loop}}$} & \textbf{$I_{w,k}$} \\
    \hline
    A--G      & $V_a$            & $I_a + K_0 I_0$                & $I_{wa,k}$ \\
    B--G      & $V_b$            & $I_b + K_0 I_0$                & $I_{wb,k}$ \\
    C--G      & $V_c$            & $I_c + K_0 I_0$                & $I_{wc,k}$ \\
    AB        & $V_a - V_b$      & $I_a - I_b$                    & $I_{wa,k} - I_{wb,k}$ \\
    BC        & $V_b - V_c$      & $I_b - I_c$                    & $I_{wb,k} - I_{wc,k}$ \\
    CA        & $V_c - V_a$      & $I_c - I_a$                    & $I_{wc,k} - I_{wa,k}$ \\
    AB--G$^{*}$ & $V_a + V_b$    & $I_a + I_b + 2K_0 I_0$         & $I_{wa,k} + I_{wb,k}$ \\
    BC--G$^{*}$ & $V_b + V_c$    & $I_b + I_c + 2K_0 I_0$         & $I_{wb,k} + I_{wc,k}$ \\
    CA--G$^{*}$ & $V_c + V_a$    & $I_c + I_a + 2K_0 I_0$         & $I_{wc,k} + I_{wa,k}$ \\
    \hline
  \end{tabular}
  \vspace{0pt}
\end{table}

As the compensation term $V_{\text{comp}}(d)$ depends explicitly on the unknown distance $d$, its evaluation requires an iterative procedure. The estimation starts by obtaining an initial value of distance  from \eqref{eq:distance_final} with $V_{\text{comp}}(d)=0$. This preliminary estimate is then substituted into \eqref{eq:compensation_voltage_generic2} to update the compensation term, yielding a new value of $V_{\text{comp}}(d)$. The updated compensation is subsequently reinserted into \eqref{eq:distance_final} to produce a refined estimate of $d$. This fixed-point iteration is repeated until the difference between successive distance estimates falls below a prescribed tolerance, ensuring convergence.

\subsection{Practical Considerations and Iterative Estimation Procedure}
\label{sec:pratical_considerations}

Table~\ref{tab:compensation} shows that $V_{\text{comp}}(d)$ depends on the inverter currents $I_{w,k}$, which are not typically obtained in a synchronous phasor-based form. Thus, when per-inverter currents are unavailable, a pragmatic approximation can use the pre-fault IED measurements as a proxy for the aggregate inverter contribution and distribute it uniformly across devices.

Assuming that: (i) all inverter generations are approximately equal during and previous the fault; and (ii) their positive-sequence injections maintain magnitude and phase angle close to pre-fault values, those assumptions can lead to an approximated form of Equation~\eqref{eq:compensation_voltage_generic2} as shown next:

\begin{equation}
  V_{\text{comp}}(d) \;=\; -\sum_{k=1}^{N} 
  \begin{cases}
 (d - d_{w,k}) \, Z_L^{(1)} \, \dfrac{I_w}{N}, & \text{if } d_{w,k} < d, \\
    0, & \text{otherwise.}
  \end{cases}
  \label{eq:compensation_voltage_pratical}
\end{equation}

\noindent where $I_w$ is a pre-fault proxy selected consistently with the loop in use. Thus, Table~\ref{tab:current_wind} enumerates these proxies for the rows of Table~\ref{tab:compensation}. It is important to observe that this approximation can estimate the fault distance $d$ using only the IED measurements.

\begin{table}[!t]
  \renewcommand{\arraystretch}{1.3}
  \centering
  \caption{Practical loop quantities for different fault types.}
  \label{tab:current_wind}
  \begin{tabular}{cccc}
    \hline
    \textbf{Fault Type} & \textbf{$V_{\text{loop}}$} & \textbf{$I_{\text{loop}}$} & \textbf{$I_w$} \\
    \hline
    A--G      & $V_a$            & $I_a + K_0 I_0$                & $I_{a}^{\text{pre}}$ \\
    B--G      & $V_b$            & $I_b + K_0 I_0$                & $I_{b}^{\text{pre}}$ \\
    C--G      & $V_c$            & $I_c + K_0 I_0$                & $I_{c}^{\text{pre}}$ \\
    AB        & $V_a - V_b$      & $I_a - I_b$                    & $I_{a}^{\text{pre}} - I_{b}^{\text{pre}}$ \\
    BC        & $V_b - V_c$      & $I_b - I_c$                    & $I_{b}^{\text{pre}} - I_{c}^{\text{pre}}$ \\
    CA        & $V_c - V_a$      & $I_c - I_a$                    & $I_{c}^{\text{pre}} - I_{a}^{\text{pre}}$ \\
    AB--G$^{*}$ & $V_a + V_b$    & $I_a + I_b + 2K_0 I_0$         & $I_{a}^{\text{pre}} + I_{b}^{\text{pre}}$ \\
    BC--G$^{*}$ & $V_b + V_c$    & $I_b + I_c + 2K_0 I_0$         & $I_{b}^{\text{pre}} + I_{c}^{\text{pre}}$ \\
    CA--G$^{*}$ & $V_c + V_a$    & $I_c + I_a + 2K_0 I_0$         & $I_{c}^{\text{pre}} + I_{a}^{\text{pre}}$ \\
    \hline
  \end{tabular}
  \vspace{-14pt}
\end{table}

\newpage
\section{Test System}

\begin{figure*}[!t]
    \centering
    \includegraphics[width=0.72\linewidth]{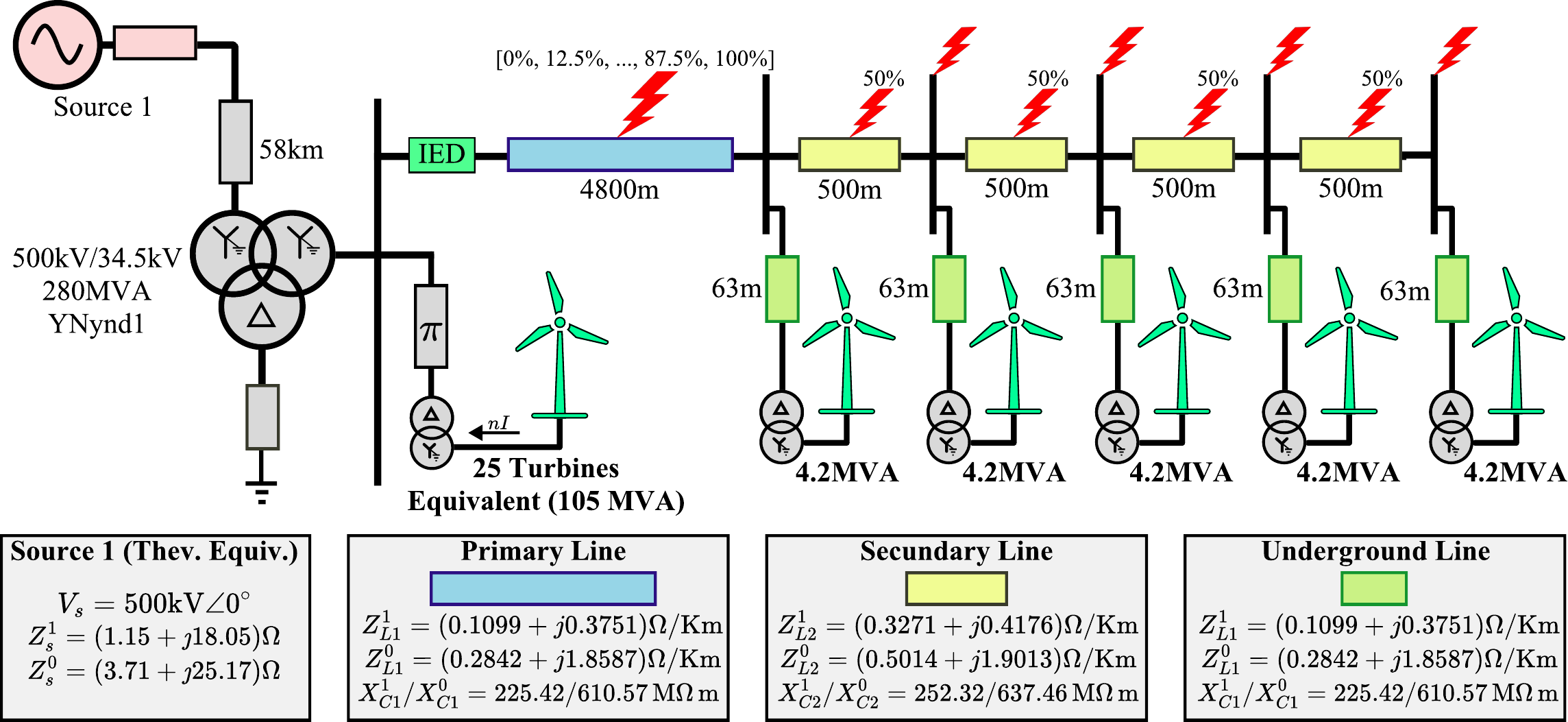}
    \vspace{-0.25cm}
    \caption{Schematic overview of the test system.}
    \label{fig:test-system}
    \vspace{-16pt}
\end{figure*}

\label{sec:test-system}

Fig.~\ref{fig:test-system} shows a realistic onshore wind-farm collector network modeled in PSCAD. The power system has 30 Type-4 wind turbine generators (WTGs), each rated at \SI{4.2}{MW}, connected via pad-mounted transformers to a collector bus. These collectors were linked through a three-winding transformer (rated at \SI{280}{MVA}) to a main collector bus, which connects to the high-voltage grid through a substation transformer.

All WTGs were modeled using full-converter technology with a grid-following (GFL) control architecture~\cite{TREMBLAY2013,MILLER2003}. The GFL inverter synchronizes with the grid via a synchronous-reference-frame phase-locked loop (PLL)~\cite{Se2000} and regulates the current through cascaded control loops. The inner current-control loop operates in the $dq$ frame and enforces the commanded current using PI controllers tuned via internal model control~\cite{Harne1998}. The outer loop manages active and reactive power set-points, with a default operation at a unity power factor during steady-state. The current magnitude was constrained to \SI{1.1}{p.u.}, and fault ride-through behavior was incorporated such that when the point of common coupling (PCC) voltage drops below \SI{0.85}{p.u.}, the control system injects positive-sequence reactive current~\cite{Dadjo2023}, as mandated by the Brazilian grid code~\cite{ONS2023}.

%To comprehensively assess the performance of the proposed methodology, a wide set of fault scenarios was simulated. The parameters and their corresponding values are summarized in Table~\ref{Tab1}. By combining all listed conditions, 17 fault locations, 10 fault types, 6 fault resistances, 3 inception angles, and 5 wind farm generation levels, a total of 15,300 unique fault cases were generated and analyzed. This extensive dataset ensures that the assessment covers a broad range of practical operating and fault conditions.

\subsection{Monte Carlo Generation of Wind-Penetration Scenarios}
\label{subsec:mc_penetration_model}

The accuracy of the proposed fault-location methodology was influenced not only by the fault location and resistance, but also by the spatial distribution of wind-power injections along the collector system. The position of each wind turbine and its local operating point affect the pre-fault current sharing and, consequently, the short-circuit level seen at the relay. In real installations, however, only the aggregated measurements at the IED bus are typically available, and the individual turbine currents are not recorded. In this work, each IBR contribution was therefore approximated as the pre-fault current at the IED divided by the number of WTGs (see Section \ref{sec:pratical_considerations}), implicitly assuming identical power output among turbines. Any dispersion in turbine output around this average may introduce additional location errors in the proposed methodology. To obtain a realistic and systematic assessment of this sensitivity, a Monte Carlo \cite{rubinstein_simulation_2016} model was adopted to generate a large set of penetration scenarios that explicitly accounts for turbine-to-turbine variability.

Field measurements reported in~\cite{WARD2023101235} indicate that, under normal operating conditions, the correlation between the power output of an individual turbine and the total farm production typically lies in the range $0.973$–$0.982$. To stress-test the proposed scheme and ensure that the simulated scenarios cover operating conditions with stronger variability than those observed in~\cite{WARD2023101235}, the penetration model was parameterized so that the correlation between each turbine and the farm-level factor remains strictly above $0.97$ for all simulated scenarios.

Let $N_\mathrm{WT}=5$ denote the number of WTGs in the wind farm. In each Monte Carlo scenario $s$, the per-unit penetration $P_i^{(s)} \in [0,1]$ of turbine $i$ was modeled as the sum of a farm-level component and a local deviation, followed by clipping to the physical interval:

\begin{equation}
  P_i^{(s)} = \mathrm{clip}\!\bigl(P_\mathrm{farm}^{(s)} + \varepsilon_i^{(s)},\,0,\,1\bigr),
  \quad i = 1,\dots,N_\mathrm{WT},
  \label{eq:penetration_model}
\end{equation}

\noindent where $P_\mathrm{farm}^{(s)}$ is the farm-level penetration common to all turbines in scenario $s$, $\varepsilon_i^{(s)}$ is a turbine-specific deviation, and $\mathrm{clip}(\cdot,0,1)$ denotes projection onto $[0,1]$. As defined in Equation~\eqref{eq:penetration_model}, the same farm-level factor drives all turbines, while $\varepsilon_i^{(s)}$ introduces controlled dispersion around this common operating point.

The farm-level penetration was modeled as uniformly distributed on the unit interval, expressed as:

\begin{equation}
  P_\mathrm{farm}^{(s)} \sim \mathcal{U}(0,1).
  \label{eq:pfarm_uniform}
\end{equation}

Thus, across scenarios, the wind farm operates between $0\%$ and $100\%$ of rated power. The deviations were independently drawn from a symmetric uniform law as:

\begin{equation}
  \varepsilon_i^{(s)} \sim \mathcal{U}(-\Delta,\Delta),
  \quad i = 1,\dots,N_\mathrm{WT},
  \label{eq:eps_uniform}
\end{equation}

\noindent where the half-width $\Delta>0$ was calibrated to enforce the desired upper bound on the correlation.

Before clipping, the variances of $P_\mathrm{farm}$ and $\varepsilon_i$ follow directly from Equations~\eqref{eq:pfarm_uniform} and~\eqref{eq:eps_uniform} as:

\begin{equation}
  \mathrm{Var}\bigl(P_\mathrm{farm}\bigr) = \frac{1}{12},
  \qquad
  \mathrm{Var}(\varepsilon_i) = \frac{\Delta^2}{3}.
  \label{eq:variances}
\end{equation}

Under the additive model implied by Equation~\eqref{eq:penetration_model}, with $P_\mathrm{farm}$ independent of $\varepsilon_i$, one obtains:

\begin{align}
  \mathrm{Var}(P_i) = \mathrm{Var}\bigl(P_\mathrm{farm}\bigr) + \mathrm{Var}(\varepsilon_i),
  \\
  \mathrm{Cov}\bigl(P_i, P_\mathrm{farm}\bigr) = \mathrm{Var}\bigl(P_\mathrm{farm}\bigr).
  \label{eq:var_cov}
\end{align}

Using the Equation~\eqref{eq:var_cov}, the correlation between turbine $i$ and the farm-level factor can be given as:

\begin{equation}
  \rho_{i,\mathrm{farm}}
  = \mathrm{Corr}(P_i, P_\mathrm{farm})
  = \sqrt{\frac{\mathrm{Var}(P_\mathrm{farm})}{
          \mathrm{Var}(P_\mathrm{farm}) + \mathrm{Var}(\varepsilon_i)}}.
  \label{eq:corr_definition}
\end{equation}

Imposing a lower bound $R_\mathrm{max}>0.97$ on this correlation and substituting Equation~\eqref{eq:variances} into~\eqref{eq:corr_definition} yields the required deviation variance:

\begin{equation}
  \mathrm{Var}(\varepsilon_i)
  = \mathrm{Var}\bigl(P_\mathrm{farm}\bigr)
    \left(\frac{1}{R_\mathrm{max}^2} - 1\right).
  \label{eq:var_eps}
\end{equation}

From Equation~\eqref{eq:var_eps}, it can be obtained the corresponding half-width:

\begin{equation}
  \Delta
  = \sqrt{3\,\mathrm{Var}(\varepsilon_i)}
  = \sqrt{
      3\,\mathrm{Var}\bigl(P_\mathrm{farm}\bigr)
      \left(\frac{1}{R_\mathrm{max}^2} - 1\right)
    }.
  \label{eq:delta_calibration}
\end{equation}

Therefore, Equation~\eqref{eq:delta_calibration} was used to select a single value of $\Delta$ that guarantees, in combination with Equation~\eqref{eq:penetration_model}, that the empirical correlation between each turbine and the farm-level penetration remains strictly above the prescribed threshold $R_\mathrm{max}$ across all simulated scenarios, while preserving strong coupling through the common factor $P_\mathrm{farm}$.

The complete sampling procedure for a single penetration scenario can be summarized as follows: (i) compute $\Delta$ from Equation~\eqref{eq:delta_calibration} using the prescribed $R_\mathrm{max}$; (ii) draw a farm-level penetration $P_\mathrm{farm}^{(s)}$ according to Equation~\eqref{eq:pfarm_uniform}; (iii) draw deviations $\varepsilon_i^{(s)}$ according to Equation~\eqref{eq:eps_uniform}; and (iv) form $P_i^{(s)}$ for all turbines using Equation~\eqref{eq:penetration_model}.

From a probabilistic standpoint, the penetration model above was embedded into the broader Monte Carlo framework used to generate the fault scenarios. Each scenario $s$ was characterized by the joint random tuple:

\begin{equation}
  \Xi^{(s)} = \bigl(P^{(s)}, F^{(s)}, D^{(s)}, R_f^{(s)}, \phi^{(s)}\bigr),
  \label{eq:xi_tuple}
\end{equation}

\noindent where $P^{(s)} = [P_1^{(s)},\dots,P_{N_\mathrm{WT}}^{(s)}]^\top$ is the penetration vector generated as described above, $F^{(s)}$ is the fault type, $D^{(s)}$ is the fault location, $R_f^{(s)}$ is the fault resistance, and $\phi^{(s)}$ is the fault inception angle. The combinations of discrete parameters considered in this work were summarized in Table~\ref{Tab1}.

\begin{table}[!ht]
    \centering
    \vspace{-0.2cm}
    \renewcommand{\arraystretch}{1.8}
    \caption{Parameters and values used to generate the scenarios.}
    \label{Tab1}
    \begin{tabular}{p{3cm} p{4.5cm}}
    \hline
    Parameters & Values  \\\hline
    Fault Location  &  17 locations shown in Fig.~\ref{fig:test-system} \\
    Fault Type  & A-G, AB, AB-G, ABC  \\
    Fault Resistance & 0$\Omega$, 5$\Omega$, 10$\Omega$, 25$\Omega$, 40$\Omega$, 50$\Omega$ \\\hline
    \end{tabular}
    \vspace{-8pt}
\end{table}

For each realization $\Xi^{(s)}$, a time-domain PSCAD simulation of the test system in Fig.~\ref{fig:test-system} was performed, and the three-phase short-circuit level and phasor quantities at the IED bus were computed and processed by the proposed algorithm.

To ensure that the set of Monte Carlo scenarios was sufficiently rich in terms of short-circuit levels, a nearest-neighbor resolution criterion was applied to the resulting fault currents. Let $I_{cc}^{(s)}$ denote the magnitude of the three-phase short-circuit current at the IED bus for scenario $s$, and let $\{I_{cc}^{(1)}, I_{cc}^{(2)}, \dots, I_{cc}^{(n)}\}$ be the first $n$ realizations. For each $n$, these values were sorted as:

\begin{equation}
  x_1^{(n)} \le x_2^{(n)} \le \cdots \le x_n^{(n)},
\end{equation}

Thus, the distance from each point to its closest neighbor in current space was computed as:

\begin{equation}
  y_i^{(n)} = \min\!\bigl( |x_i^{(n)} - x_{i-1}^{(n)}|,\,
                           |x_{i+1}^{(n)} - x_i^{(n)}| \bigr),
  \quad i=1,\dots,n,
\end{equation}

\indent with the convention that $x_0^{(n)}$ and $x_{n+1}^{(n)}$ were omitted at the boundaries. The $p$-th percentile of these nearest-neighbor distances, $\varepsilon_{p}(n)
  = \mathrm{Percentile}_p\bigl(\{y_i^{(n)}\}_{i=1}^n\bigr)$, provides a measure of how densely the short-circuit current space was sampled. In this work, $p=99$ was adopted, so that $\varepsilon_{99}(n)$ was the distance such that $99\%$ of the simulated short-circuit levels have another scenario within $\varepsilon_{99}(n)$ amperes. A tolerance $\Delta I_{cc,\mathrm{tol}}$ was then specified (here, $\Delta I_{cc,\mathrm{tol}} = 10~\mathrm{A}$), and the Monte Carlo process was deemed converged when $\varepsilon_{99}(n) \le \Delta I_{cc,\mathrm{tol}}$.

\begin{figure}[!ht]
    \centering
    \includegraphics[width=0.9\linewidth]{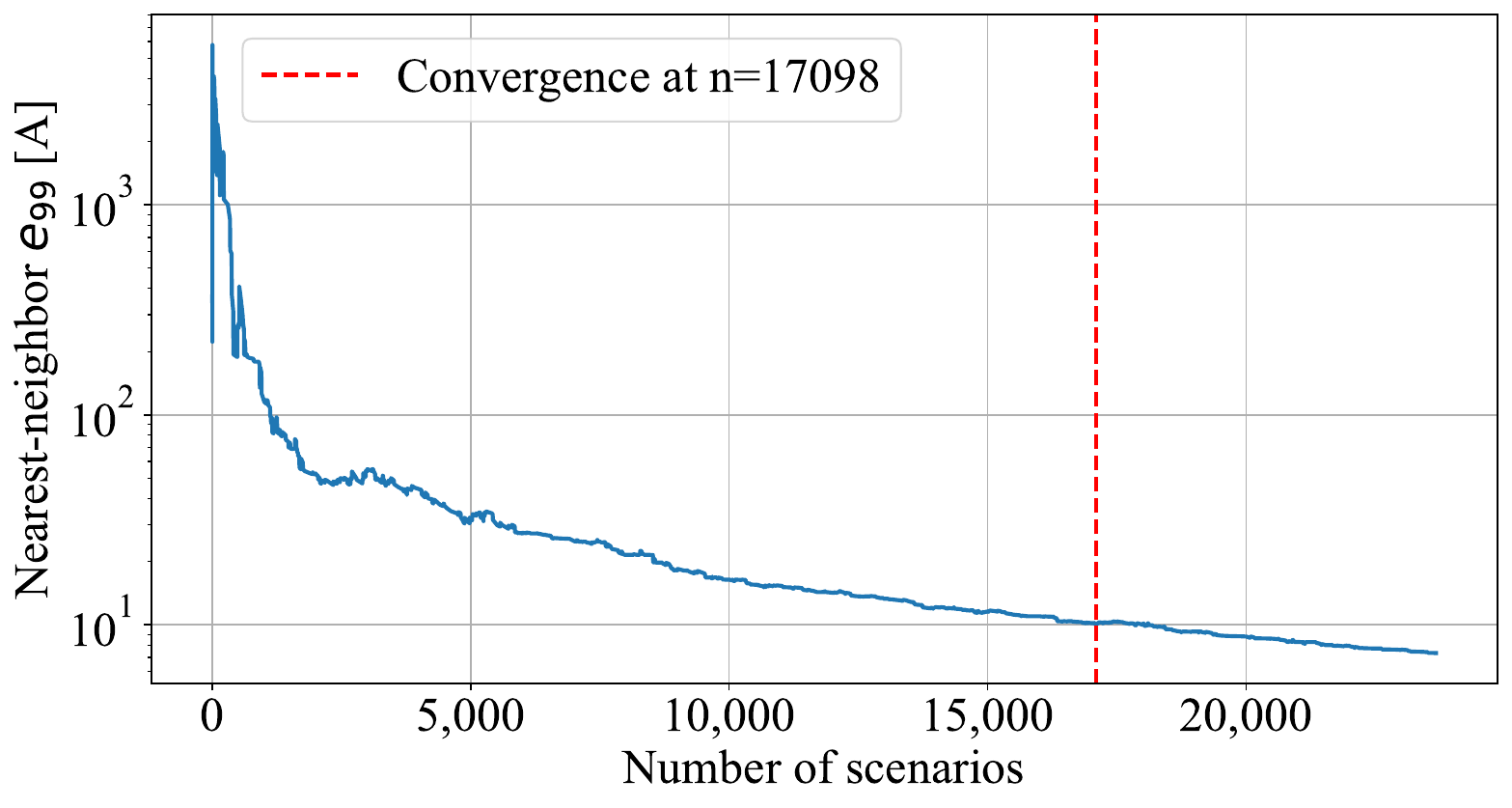}
    \vspace{-0.5cm}
    \caption{Nearest-neighbor Monte Carlo convergence of the short-circuit current at the IED bus.}
    \label{fig:mc-convergence}
    
\end{figure}

Fig.~\ref{fig:mc-convergence} shows the evolution of $\varepsilon_{99}(n)$ as a function of the number of scenarios, together with the convergence threshold. As the number of simulations increases, the nearest-neighbor distance decreases approximately monotonically, indicating progressively finer resolution of the short-circuit current distribution. The resolution criterion was satisfied only when the Monte Carlo set approaches its final size, implying that additional simulations beyond this point would not introduce materially new short-circuit levels, but only refine already-sampled regions.

\begin{figure}[!t]
    \centering
    \includegraphics[width=0.9\linewidth]{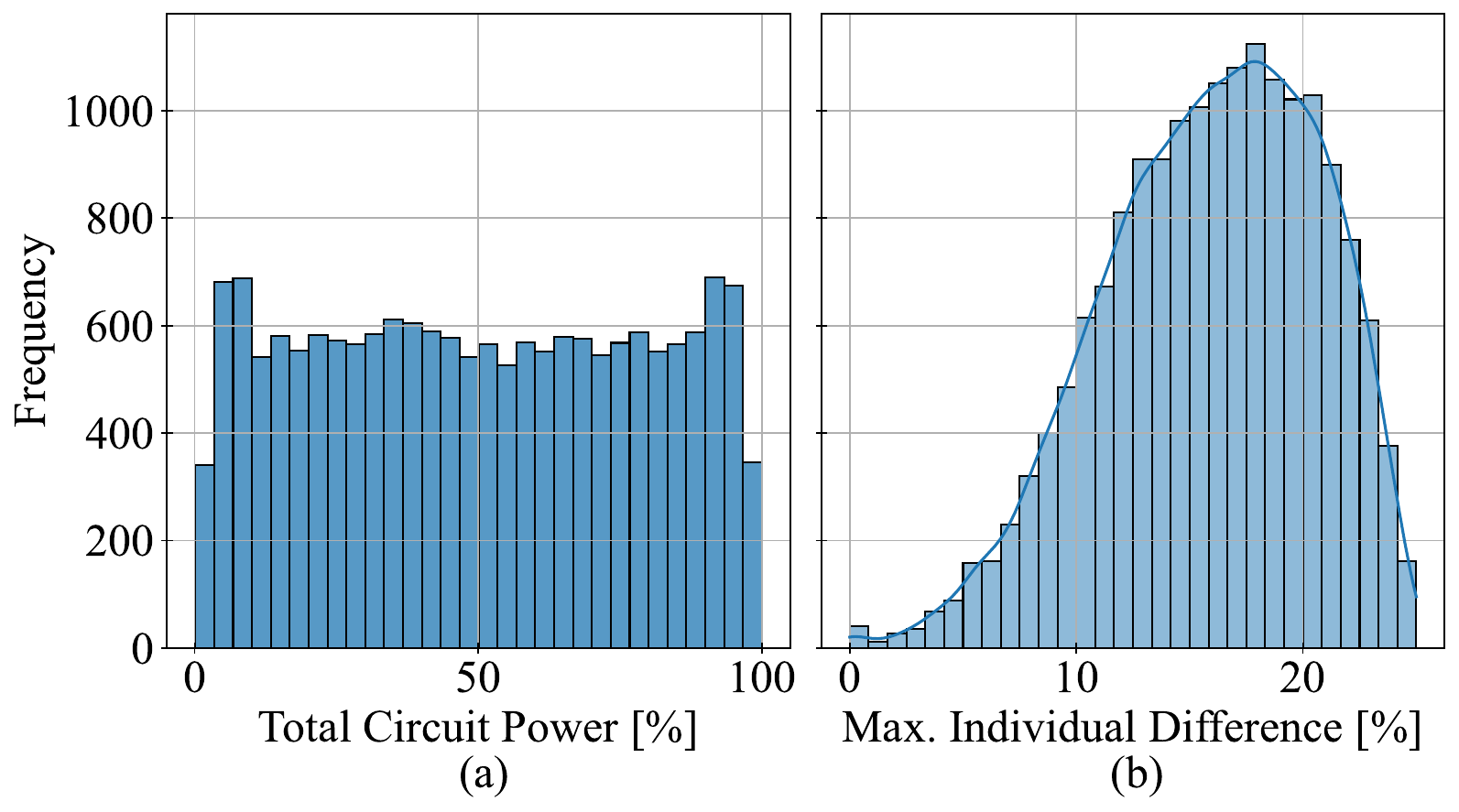}
    \vspace{-0.4cm}
    \caption{Monte Carlo statistics of wind penetration: (a) distribution of the total circuit power across all scenarios; (b) distribution of the maximum turbine-to-turbine penetration difference.}
    \label{fig:mc-power-distribution}
    \vspace{-18pt}
\end{figure}

Based on the converged Monte Carlo set, the resulting distributions of the total wind-farm penetration and of the maximum turbine-to-turbine difference across all Monte Carlo realizations were illustrated in Fig.~\ref{fig:mc-power-distribution}. The final Monte Carlo set encompass the 17,098 fault cases analyzed in this study.

\section{Assessment of the Methodology}
\label{sec:results}

To assess the proposed methodology, its performance was benchmarked against state-of-the-art one-terminal phasor-based fault-location techniques. For each fault category, the most accurate method reported in the literature was selected to ensure a fair and representative comparison. All benchmark locators were implemented using the general compensated formulation in Equation~\eqref{eq:distance_final}, with the loop-dependent voltage and current quantities defined according to the corresponding fault type.

\begin{figure}[!ht]
  \centering
  \includegraphics[width=1\linewidth]{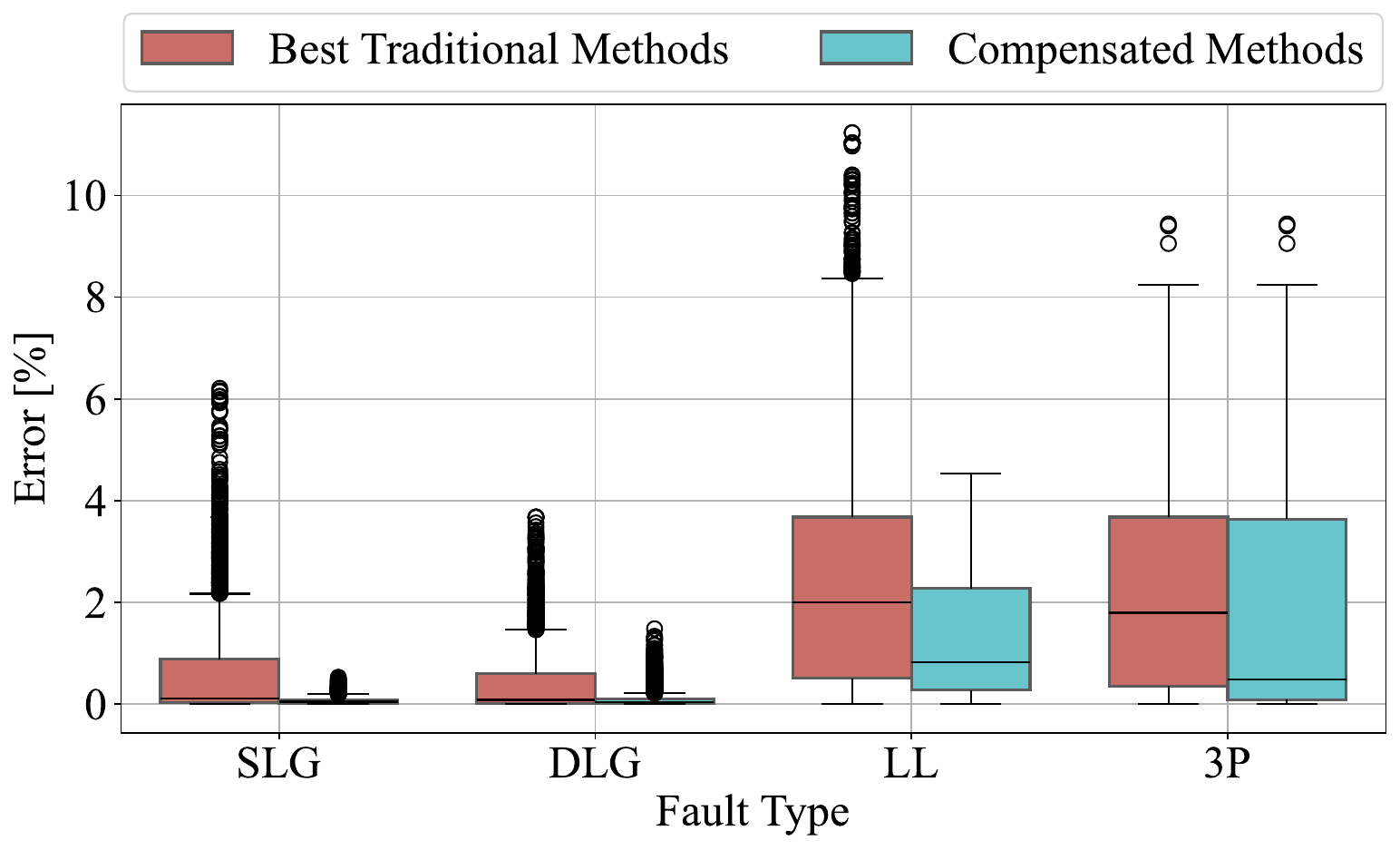}
  \vspace{-0.8cm}
  \caption{Fault location error comparison between the most promising locator methods with and without the compensation voltage.}\label{fig:boxplot_fault_type}
  \vspace{-18pt}
\end{figure}

The evaluation employed the simulation scenarios described in Section~\ref{sec:test-system}, enabling a consistent comparison of fault-location accuracy between the best conventional approaches and the proposed method. Following the recommendations in \cite{DAVI2025MULTI}, the TAKZ \cite{TAKAGI1982} algorithm was adopted for SLG faults, the enhanced TAKZ\_New \cite{MOISESLOC1} formulation for double-line-to-ground (DLG) faults, the TAKN \cite{SEL2018} method for line-to-line (LL) faults, and the Reactance \cite{CAPAR2014} approach for three-phase (3P) faults. The overall performance across these four major fault classes is summarized in Fig.~\ref{fig:boxplot_fault_type}.

The boxplots show a contrast in fault location errors when compensation was applied. Notably, SLG and DLG faults exhibit exceptional improvement, with error distributions concentrated near zero and very narrow interquartile ranges. This enhanced accuracy stems from the fact that both fault types rely on the zero-sequence current component to estimate the fault current, a quantity unaffected by inverter injections due to the $\Delta$-Y configuration of the inverter step-up transformers. In contrast, LL faults use the negative-sequence component, which can be distorted by fault ride-through controls that inject negative-sequence current when the voltage drops below a threshold \cite{8506376}. For 3P faults, which depend solely on the positive-sequence component, the situation worsens because the injected current was significantly altered even under normal conditions, making the benefits of compensation less significant, although still substantial.

Complementing this qualitative insight, Table~\ref{tab:error_stats_ibrloc} quantifies the average location error for each fault type, with and without the proposed compensation.  These results corroborate the trends observed in Fig.~\ref{fig:boxplot_fault_type}, i.e., average errors for SLG and DLG faults drop from 0.584\% and 0.386\% to 0.065\% and 0.095\%, respectively (improvements that exceed 75\%). LL and 3P faults also benefit, with error reductions of approximately 46\% and 16\%, respectively. These findings confirm that the compensation term significantly mitigates the IBR-induced deviations, especially for faults that are not dominated by positive-sequence contributions.
\vspace{-10pt}
\begin{table}[!ht]
  \centering
  \renewcommand{\arraystretch}{1.3}
  \caption{Average fault location error for the most promising locator methods with and without the compensation voltage.}
  \label{tab:error_stats_ibrloc}
  \begin{tabular}{lcccc}
  \hline
  Method                         & SLG       & DLG       & LL        & 3P        \\
  \hline
  Impedance \cite{ZIEGLER2011}        & 24.659\% & 29.601\% & 27.649\% & 24.462\% \\
  Reactance \cite{CAPAR2014}         & 27.739\% & 12.269\% &  9.124\% &  2.199\% \\
  TAKS \cite{TAKAGI1982}                    &  3.592\% & 18.772\% &  6.871\% & 11.028\% \\
  TAKN \cite{SEL2018}                       &   --     &   --     &  2.455\% &   --     \\
  TAKZ \cite{TAKAGI1982}                    &  0.584\% & 39.259\% &   --     &   --     \\
  TAKZ\_New \cite{MOISESLOC1}               &   --     &  0.386\% &   --     &   --     \\
  \textbf{Proposed Method}                  &  \textbf{0.065\%} & \textbf{0.095\%} & \textbf{1.310\%} & \textbf{1.838\%} \\
  \hline
  \textbf{Improvement} & 
    \textbf{88.95\%} & 
    \textbf{75.47\%} & 
    \textbf{46.66\%} & 
    \textbf{16.41\%} \\
  \hline
  \end{tabular}
\end{table}

Further demonstrating the robustness of the proposed methodology, Table~\ref{tab:maxerror_stats_ibrloc2} reports the maximum fault location errors observed across the full simulation set. The results show that, even under the most adverse operating conditions, the proposed approach consistently outperforms the traditional techniques. The improvement was most pronounced for SLG faults, where the worst-case error was reduced from 6.208\% to 0.529\%. DLG and LL faults exhibit similarly strong gains, with reductions exceeding 59\% in their respective maximum errors. In contrast, no improvement was observed for the 3P fault category. This outcome was attributed to the fact that the peak error in this case originates from a fault on the primary line, where the proposed formulation does not yield additional corrective benefit.

\begin{table}[!ht]
  \centering
  \renewcommand{\arraystretch}{1.3}
  \caption{Max fault location error for the most promising locator methods with and without the compensation voltage.}
  \label{tab:maxerror_stats_ibrloc2}
  \begin{tabular}{lcccc}
    \hline
    Method                         & SLG       & DLG       & LL        & 3P        \\
    \hline
    Impedance \cite{ZIEGLER2011}   & 100.000\% & 100.000\% & 100.000\% & 100.000\% \\
    Reactance \cite{CAPAR2014}     &  97.903\% & 100.000\% &  89.973\% &   9.439\% \\
    TAKS \cite{TAKAGI1982}                &  30.019\% & 100.000\% &  61.302\% &  69.447\% \\
    TAKN \cite{SEL2018}                   &    --     &    --     &  11.234\% &    --     \\
    TAKZ \cite{TAKAGI1982}                &   6.208\% & 100.000\% &    --     &    --     \\
    TAKZ\_New \cite{MOISESLOC1}           &    --     &   3.676\% &    --     &    --     \\
    \textbf{Proposed Method}              & \textbf{0.529\%} & \textbf{1.482\%} & \textbf{4.537\%} & \textbf{9.439\%} \\
    \hline
    \textbf{Improvement} & 
      \textbf{91.48\%} & 
      \textbf{59.69\%} & 
      \textbf{59.61\%} & 
      \textbf{0.00\%} \\
    \hline
  \end{tabular}
  \vspace{-12pt}
\end{table}

\begin{figure}[!ht]
  \centering
  \includegraphics[width=\linewidth]{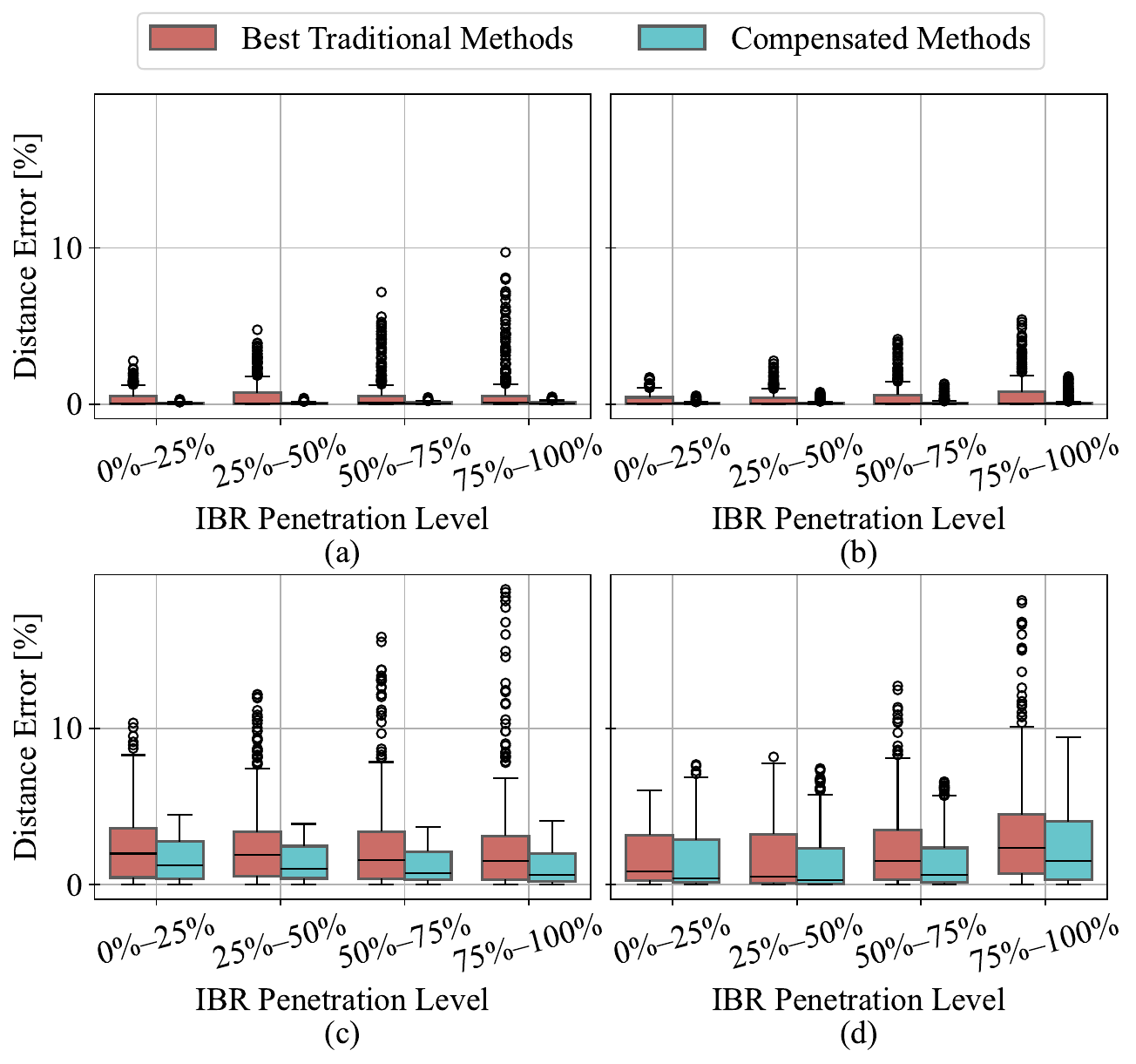}
  \vspace{-0.8cm}
  \caption{Fault location error comparison between the most promising locator methods with and without the compensation voltage for the fault types: (a) SLG, (b) DLG, (c) LL, and (d) 3P.}\label{fig:boxplot_fault_penetration}
  \vspace{-6pt}
\end{figure}

The influence of wind power penetration on fault location performance was illustrated in Fig.~\ref{fig:boxplot_fault_penetration}. Without compensation, the error increases markedly with higher levels of wind generation, reflecting the growing impact of IBR current injections on the voltage and current phasors measured at the IED. In contrast, the compensation methodology neutralizes this effect, yielding nearly constant error distributions across all generation levels. This invariance to wind penetration levels underscores the method’s scalability and adaptability to evolving system conditions with higher renewable integration.

\begin{figure}[!ht]
    \centering
    \includegraphics[width=1\linewidth]{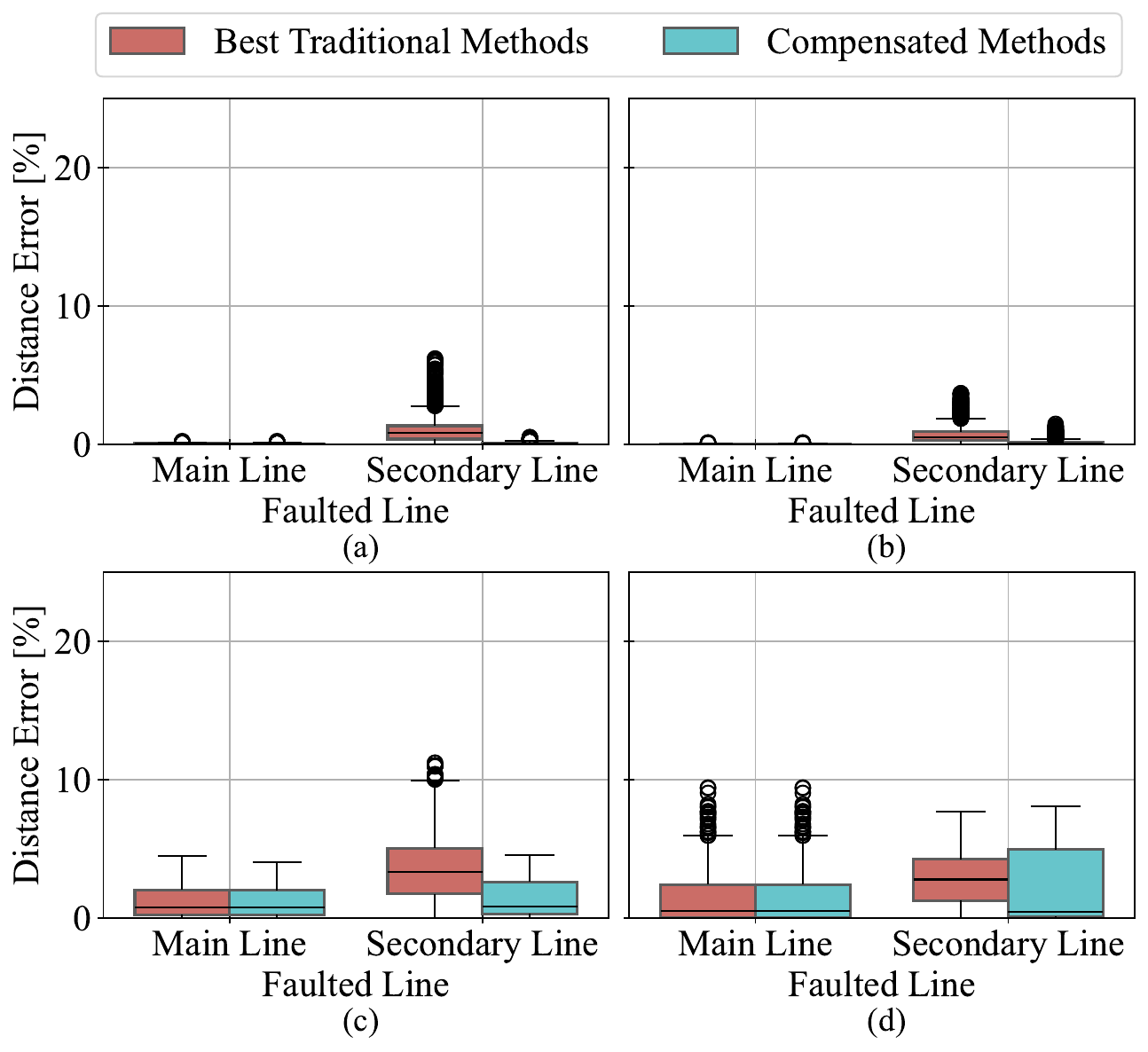}
    \vspace{-0.8cm}
    \caption{Fault location error comparison between the primary and secondary lines for the fault types: (a) SLG, (b) DLG, (c) LL, and (d) 3P.}\label{fig:boxplot_fault_dist}
    \vspace{-12pt}
\end{figure}

Finally, Fig.~\ref{fig:boxplot_fault_dist} compares the error distribution between faults occurring in the primary line (upstream of any inverter) and the secondary line, where IBR injections are present between the IED and the fault. As expected, in the primary line, compensated and uncompensated errors were nearly identical as no IBRs intervene in the loop. However, for the secondary line, the uncompensated error was significantly larger, reflecting the deviations introduced by unaccounted inverter injections. The compensated results in the secondary line closely match those of the primary line, demonstrating that the methodology effectively attenuates the influence of IBRs when their position lies between the IED and the fault.

Taken together, the results shown in these figures and tables demonstrate that the proposed compensation methodology offers substantial and consistent gains in fault location accuracy across fault types, system conditions, and topological configurations. Its ability to mitigate the IBR-induced deviations, particularly in complex wind farm collector systems, suggests that it can serve as a robust enhancement to conventional one-terminal phasor-based locators.

\section{Conclusion}
\label{sec:conclusion}

This paper highlighted a fundamental limitation in traditional one-terminal phasor-based fault location methods when applied to wind farm collector networks, namely their reduced accuracy in the presence of IBRs located downstream from the fault point. To mitigate this effect, a simple compensation framework was proposed to correct loop quantities distorted by IBR current injections, preserving the structural simplicity of classical schemes while enhancing their accuracy under high IBR penetration. The methodology was assessed in PSCAD using a realistic onshore wind farm with 27 full-converter turbines and more than 15{,}000 fault scenarios spanning different fault types, resistances, inception angles, locations, and wind penetration levels.

Quantitative results showed substantial reductions in both average and maximum fault-location errors, exceeding 75\% for SLG and DLG faults, and significant improvements for LL and 3P faults. The proposed approach also reduced the sensitivity of the locator to variations in wind penetration and helped equalize performance along the feeder, irrespective of IBR placement. Overall, the compensation framework provides a low-complexity, method-agnostic enhancement that restores the effectiveness of phasor-based fault-location schemes in wind-farm collector systems with high IBR integration.

\bibliographystyle{IEEEtran}
\bibliography{refs.bib}
\vspace{-8pt}
\section*{Acknowledgments}

The authors would like to acknowledge the support of the National Council for Scientific and Technological Development under Grant No.~307982/2022-0. They also gratefully acknowledge the Research Centre for Greenhouse Gas Innovation at the University of São Paulo, funded by the São Paulo Research Foundation under Grant No.~2020/15230-5 and by TotalEnergies, as well as the strategic support provided by the Brazilian National Oil, Natural Gas and Biofuels Agency through the R\&D levy regulation.
\vspace{-24pt}
% \vspace{-48pt}

\vfill

\end{document}